\documentclass[journal=jacsat,manuscript=article]{achemso}
\usepackage{todonotes}
\usepackage{graphicx}
\usepackage{siunitx}
\usepackage{amsmath}
\graphicspath{ {./img/} }
\usepackage[table]{xcolor}
\usepackage[T1]{fontenc} 

\usepackage{makecell}
\usepackage{booktabs}
\usepackage{multirow}
\usepackage{orcidlink}
\usepackage{hyperref}
\definecolor{niceblue}{RGB}{0,40,180}  
\hypersetup{
    colorlinks=true,
    linkcolor=niceblue,
    urlcolor=niceblue,
    citecolor=niceblue
}

\usepackage{amssymb}
\usepackage[highlightmarkup=uwave]{changes}
\title{Low-Noise Nanoscale Vortex Sensor for Out-of-Plane Magnetic Field Detection}

\author{Ajay Jha\orcidlink{0000-0001-5738-8649}} 
\email{ajay-kumar.jha@cea.fr}
\affiliation{University Grenoble Alpes, CEA, CNRS, Grenoble-INP, Spintec, 38000 Grenoble, France}

\author{Alvaro Palomino\orcidlink{0000-0003-0313-0110}}
\affiliation{University Grenoble Alpes, CEA, CNRS, Grenoble-INP, Spintec, 38000 Grenoble, France}

\author{Stéphane Auffret\orcidlink{0009-0006-5008-227X}}
\affiliation{University Grenoble Alpes, CEA, CNRS, Grenoble-INP, Spintec, 38000 Grenoble, France}

\author{Hélène Béa\orcidlink{0000-0002-3762-4795}}
\affiliation{University Grenoble Alpes, CEA, CNRS, Grenoble-INP, Spintec, 38000 Grenoble, France}
\alsoaffiliation{Institut Universitaire de France (IUF), 75000 Paris, France}

\author{Ricardo C. Sousa\orcidlink{0000-0001-8903-3359}}
\affiliation{University Grenoble Alpes, CEA, CNRS, Grenoble-INP, Spintec, 38000 Grenoble, France}

\author{Liliana D. Buda-Prejbeanu\orcidlink{0000-0002-6105-151X}}
\affiliation{University Grenoble Alpes, CEA, CNRS, Grenoble-INP, Spintec, 38000 Grenoble, France}

\author{Bernard Dieny\orcidlink{0000-0002-0575-5301}} 
\email{bernard.dieny@cea.fr}
\affiliation{University Grenoble Alpes, CEA, CNRS, Grenoble-INP, Spintec, 38000 Grenoble, France}
\begin{document}

\begin{abstract}
{\setlength{\parindent}{0cm}
This study investigates a vortex sensor based on a nanoscale (sub-100 nm)  magnetic tunnel junction (MTJ) with a strong shape anisotropy, designed for sensitivity to the out-of-plane magnetic field component ($H_z$). The sensor comprises a free layer with a vortex configuration and a perpendicularly magnetized reference layer, which provides a reproducible and linear response when excited by a perpendicular magnetic field. Experimental measurements and micromagnetic simulations were combined to systematically assess the influence of structural parameters, specifically aspect ratio and defect landscape, on key sensor performance metrics, including dynamic range, sensitivity, and detectivity. The out-of-plane vortex sensor demonstrates a significantly improved dynamic range exceeding 200 mT, compared to the 40–80 mT typical of conventional in-plane vortex sensors. Frequency-dependent noise measurements reveal that the sensor exhibits low intrinsic noise, along with improved detectivity and resolution. This performance is ascribed to the field-dependent expansion and contraction of the vortex core, which reduces Barkhausen-type noise caused by defect-induced pinning potentials. Moreover, the sub-100\,nm lateral dimensions of the sensor enable scalable array integration, providing further enhancements in noise and detectivity through collective averaging. These results underscore the potential of this sensor architecture for advanced magnetic field sensing applications requiring a wide dynamic range and high measurement accuracy at the same time.\\


}
\end{abstract}
\maketitle
\section{Introduction}
\label{sec:introduction}

Magnetic field sensors are essential elements in a wide range of technological domains, including biomedical diagnostics \cite{kwon2024anisotropy, lin2017magnetic, murzin2020ultrasensitive, freitas2012spintronic}, automotive electronics~\cite{silva2015linearization, babatain2021acceleration}, industrial automation \cite{silva2015linearization}, geophysical instrumentation~\cite{zuo2020miniaturized, manceau2023large} and  consumer electronics such as read head of hard disk drives (HDD)~\cite{shiroishi2009future, gallagher2006development}. Among the widely employed sensor technologies are Hall effect sensors~\cite{ramsden2011hall}, anisotropic magnetoresistance (AMR) sensors~\cite{dietmayer2001magnetische}, and fluxgate magnetometers~\cite{auster2008themis, ripka2003advances}. These conventional technologies are favored because of their structural simplicity, cost-effective manufacturing, and well-established knowledge regarding their integration with read-out electronics~\cite{shiroishi2009future}.
However, these sensors are inherently limited in sensitivity, with the exception of fluxgate magnetometers, which provide exceptionally high sensitivity~\cite{auster2008themis, ripka2003advances}. Nevertheless, all magnetic sensors, including fluxgates, continue to suffer limitation in spatial resolution, bandwidth, and/or power efficiency. These issues become particularly critical in applications that demand device miniaturization, low-power operation, or high-frequency responsiveness. These challenges have prompted the investigation of alternative magnetic sensing solutions with improved performance.

The discovery of magnetoresistive sensors based on spintronic phenomena~\cite{freitas2007magnetoresistive}, particularly magnetic tunnel junctions (MTJs) and spin-valve structures, has emerged as highly promising candidates~\cite{dieny1991giant, heim2002design, freitas2000spin}. MTJ sensors, exploiting the tunneling magnetoresistance (TMR) effect~\cite{julliere1975tunneling, parkin2004giant}, offer several distinct advantages: exceptional field sensitivity, compact form factor, low power consumption, and full compatibility with complementary metal oxide-semiconductor (CMOS) processes~\cite{nakano2025tunnel, freitas2016spintronic, freitas2000spin}. These features make them highly suitable for monolithic integration into advanced microelectronic platforms~\cite{zuo2020miniaturized} and ideal for precision applications such as current sensing, rotational speed detection, and angular position tracking in automotive and industrial systems~\cite{silva2015linearization, babatain2021acceleration}. However, conventional MTJ sensors are constrained by an inherent trade-off between sensitivity, linearity, and magnetic hysteresis, which limits their performance—particularly in dynamic or high-resolution sensing applications.

To address these limitations, significant research has been directed toward integrating magnetic vortex states into MTJ sensors~\cite{cowburn1999single, guslienko2001magnetization}. A magnetic vortex is a stable magnetization configuration commonly observed in micrometer-scale soft ferromagnetic disks. It arises in the regime where magnetostatic energy dominates over exchange energy and is characterized by an in-plane curling of magnetization surrounding a nanoscale out-of-plane core~\cite{fernandez1998magnetic, guslienko2001magnetization, wurft2019evolution}. This configuration is energetically favorable, yielding low stray fields and low hysteresis in the vortex state. When such a vortex state is embedded in the free layer of an MTJ, it enables a highly linear, and reproducible response to external in-plane magnetic stimuli~\cite{suess2018topologically, he2020nonhysteretic, weitensfelder2018comparison}. The lateral displacement of the vortex core under the applied in-plane field translates into a measurable TMR signal, providing a robust mechanism for field detection.

\begin{figure*}[!ht]
\centering
     \includegraphics[width=\linewidth]{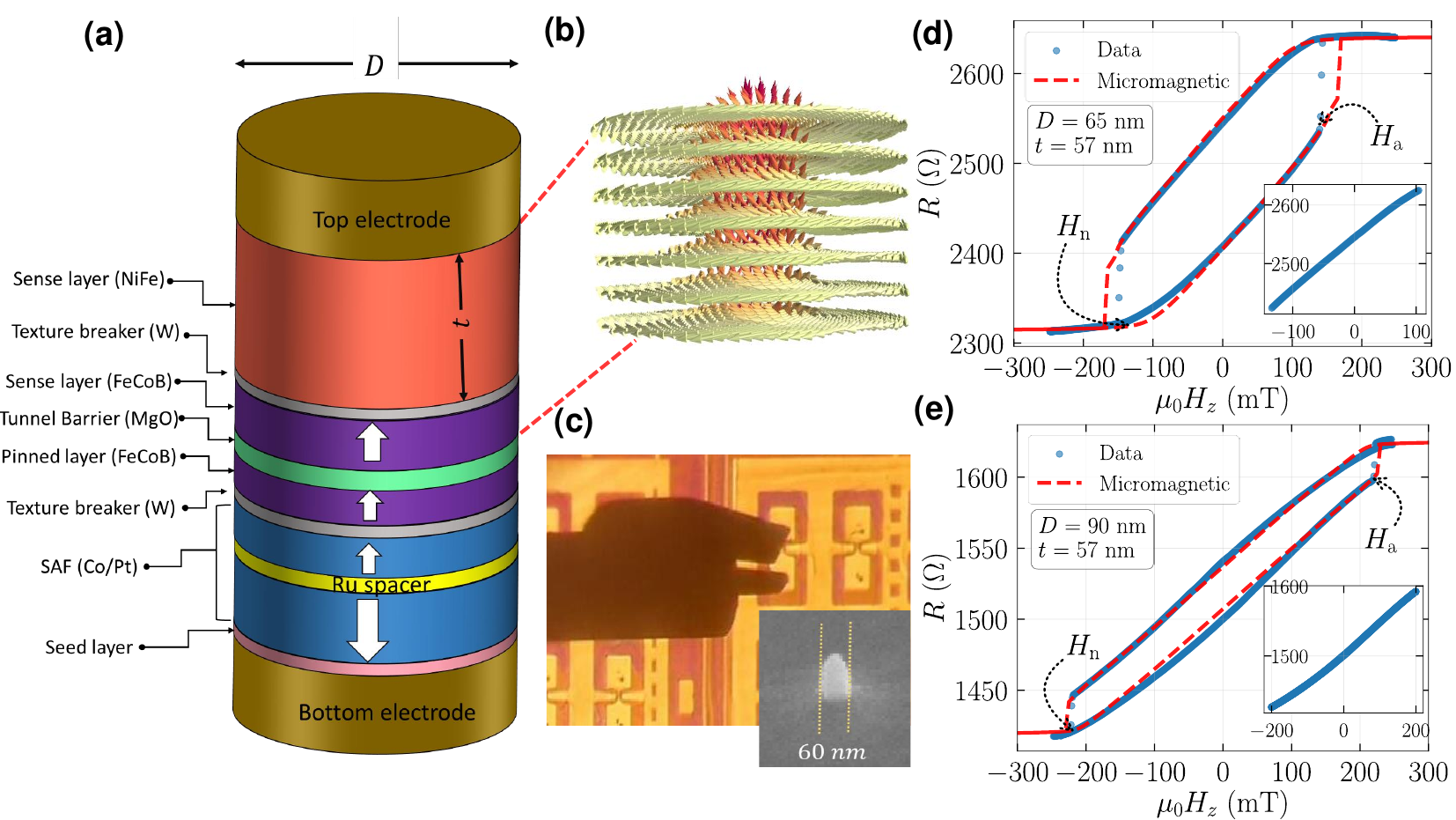}
    \captionsetup{singlelinecheck=false,labelformat=simple, width=\linewidth}
	\caption{  (a) Schematic illustration of the vortex-based MTJ sensor, highlighting key components such as the SAF layers, PMA-reference/pinned layer, tunnel barrier, and the composite sensing layer. The sensing layer consists of a combination of PMA FeCoB (1.4 nm) and NiFe (57 nm) layers to support vortex formation. (b) Schematic of the magnetic vortex configuration in the sensing layers. Within the core region, the magnetization is oriented perpendicular to the plane, while outside the core, the magnetization exhibits an in-plane curling configuration.
    (c) Optical micrograph of the device under test and SEM image of the nanopillar.
    (d-e) Experimental transfer curves (blue dots) and corresponding micromagnetic simulations (red dashed lines) for devices with diameters \( D = 65~\text{nm} \) and \( D = 90~\text{nm} \), respectively. Both devices feature sensing layers with a thickness of approximately 57 nm.  The nucleation field \( H_\mathrm{n} \) and annihilation field \( H_\mathrm{a} \) are indicated for each device. The insets show the experimentally obtained reversible linear response when the magnetic field is constrained below the critical thresholds \( H_\mathrm{n} \) and \( H_\mathrm{a} \).
    }
	\label{fig:figure_1}
\end{figure*}

Vortex-based MTJ sensors offer several compelling advantages over traditional magnetoresistive designs. Their closed magnetic flux structure inherently suppresses dipolar field interactions with the reference layer, improving stability and compact sensor array designs without interference between neighboring elements.
Moreover, their symmetric geometry enables scalable and low-power operation over a wide linear working field range. Such characteristics make them particularly attractive in advance field sensing application~\cite{suess2018topologically, he2020nonhysteretic}.

In conventional vortex-based MTJ sensors, which typically have diameters around 1-5\,\(\mu\)m and a low aspect ratio (\( t/D \ll 1 \), where \( D \) is the diameter and \( t \) is the thickness of the sensing layer), the sensing layer stabilizes into a vortex magnetic state. This configuration is predominantly sensitive to in-plane magnetic fields, as illustrated in Figure~S1 of the Supporting Information. Upon application of an in-plane magnetic field, the vortex core undergoes lateral displacement, perpendicular to the field direction, within the plane of the magnetic layer. This motion, governed by the rigid vortex model~\cite{guslienko2001magnetization, thiele1973steady}, leads to a change in the local magnetization orientation relative to the pinned reference layer, modulating the TMR and generating a corresponding electrical output.
The sensor typically exhibits a quasi-linear transfer characteristic within a defined magnetic field range, rendering it highly suitable for field sensing applications. However, the output signal often exhibits discrete resistance jumps~\cite{jotta2021spin, guslienko2001field}. These discontinuities are typically caused by the existence of intrinsic material defects, such as grain boundaries in polycrystalline samples. Another contributing factor is anisotropy fluctuations at the interface between the tunnel barrier (MgO) and the sensing layer. In addition, extrinsic fabrication imperfections may further exacerbate these effects (see Figure~S1 in the Supporting Information).
These variations lead to Barkhausen noise which arises from pinning and depinning of the vortex core as it overcomes local energy barriers~\cite{burgess2013quantitative, barkhausen1919zwei}.

To address this issue, we propose an alternative class of vortex-based MTJ sensors that are sensitive to out-of-plane magnetic fields~\cite{palomino_magnetorestistive_2023}. This approach employs device geometries with larger aspect ratios (\( t/D \sim 0.4\text{--}1 \)) and significantly smaller diameters (60--150\,nm), in contrast to conventional in-plane vortex sensors, which typically have low aspect ratios (\( t/D \ll 1 \)) and diameters around 1-5 \,\(\mu\)m. In such configurations, the vortex core contracts or expands in response to variations in the perpendicular magnetic field (\( H_z \)), leading to a gradual and nearly linear modulation of the out-of-plane magnetization component (\( m_z \)). This modulation, in combination with a reference layer exhibiting strong perpendicular magnetic anisotropy, results in a linear TMR response over the applied field range. The linear response arising from the expansion and contraction of the vortex core suppresses Barkhausen-type noise, as this mechanism avoids lateral core motion, which is susceptible to local pinning potentials.

Here, we present a comprehensive investigation of the proposed sensor architecture through both experimental measurements and micromagnetic simulations. The sensor exhibits a wide dynamic range exceeding \( 200\,\mathrm{mT} \), competitive detectivity, and superior resolution, outperforming previously reported single-element MTJ-based vortex sensors
~\cite{he2020nonhysteretic, suess2018topologically, khan2021magnetic, ripka2010advances, weitensfelder2018comparison}. With a lateral dimension below 100\,nm, the sensor maintains an ultra-compact footprint, and its performance can be further enhanced through array-based integration. These attributes underscore the scalability and application potential of the proposed design for next-generation magnetic field sensing technologies, where low noise, wide dynamic range, and high integration density are essential.

\section*{Results and Discussion}
\subsection*{Experimental vs simulation}

To investigate the properties of vortex-based MTJ sensors, multilayer thin-film stacks were deposited onto 100~mm Si wafers using an ultra-high vacuum (UHV) magnetron sputtering system. Detailed descriptions of the deposition process and subsequent nanofabrication steps are provided in Section~\hyperlink{sec:methods}{Methods}. The MTJ stack (\autoref{fig:figure_1}a) comprises a Co/Pt-based synthetic antiferromagnetic (SAF) structure, exchange-coupled to an adjacent FeCoB layer. This coupling induces strong perpendicular magnetic anisotropy (PMA) in the FeCoB, enabling it to function as the reference layer with its magnetization oriented along the out-of-plane (\( z \)-axis) direction.
The free sensing layer consists of a composite structure: a thin FeCoB (1.4 nm) layer deposited directly on the MgO tunnel barrier, followed by a thick NiFe layer with a  thickness of $t \sim 57$ nm. The FeCoB layers at the MgO interfaces enhance the crystalline quality of the barrier and improve the TMR, both critical factors for high-performance sensor applications. The large thickness of the NiFe layer promotes the stabilization of a vortex state in the sense layers (\autoref{fig:figure_1}b), a configuration that becomes energetically favorable when magnetostatic energy outweighs exchange energy. Through exchange coupling across the W layer, the vortex state is also imprinted in the FeCoB layer, which dominates the electrical properties of the TMR sensor. 

Electrical measurements were performed on devices with diameters ranging from 60 to 100 nm. \autoref{fig:figure_1}d and \autoref{fig:figure_1}e show the transfer curves of the measured resistance versus magnetic field ($R$–$H$) for devices with diameters of 65 nm and 90 nm, respectively. These curves clearly illustrate the nucleation of the vortex core when the applied magnetic field drops below a critical threshold, known as the nucleation field ($H_\mathrm{n}$). As the magnetic field is further varied, the size of the vortex core gradually changes until the field reaches the annihilation point ($H_\mathrm{a}$) at which the vortex collapses and the magnetization enters a saturated state. A broad quasi-linear response is observed between $H_\mathrm{n}$ and $H_\mathrm{a}$ representing the effective operating range of the sensor.

To elucidate the magnetization evolution, micromagnetic simulations were performed for both devices. The simulations incorporated structural disorder and local pinning potentials intrinsic to sputtered films, modeled using a Voronoi tessellation approach as detailed in Section~\hyperlink{sec:methods}{Methods}. The simulated transfer curves exhibited excellent agreement with the experimental data, accurately reproducing all key features of the measured responses (\autoref{fig:figure_1}d–e).
The overall magnetization response exhibits two quasi-linear parallel branches and abrupt transition to saturation corresponding to the annihilation of the vortex configuration. The two branches are associated to two different orientations of the vortex core magnetization. Indeed, when increasing the field from negative saturation, the vortex core nucleates in the negative field direction while when decreasing the field from positive saturation, it nucleates in the positive field direction. This yields the observed hysteresis of the major loop. However, if the field range is maintained below the annihilation field threshold, a reversible quasi-linear response is obtained with an operating range exceeding 200 mT (\autoref{fig:figure_1}d–e). This quasi-linear region is particularly suitable for sensor applications requiring reproducible output within a limited magnetic field range. Furthermore, to suppress the existence of two separate branches, several approaches can be employed to force the vortex core nucleation to always occur with the same polarity. One method involves shifting the vortex annihilation field to higher values by applying an external offset magnetic field~\cite{silva2015linearization}. Alternatively, a comparable effect can be realized either through exchange-bias coupling between an antiferromagnetic layer and the ferromagnetic sensing layer~\cite{stearrett2012influence, chappert2007emergence, palomino_magnetorestistive_2023}, or by tailoring the geometry of the sensing layer.
\begin{figure*}[!htb]
\centering
    \includegraphics[width=\linewidth]{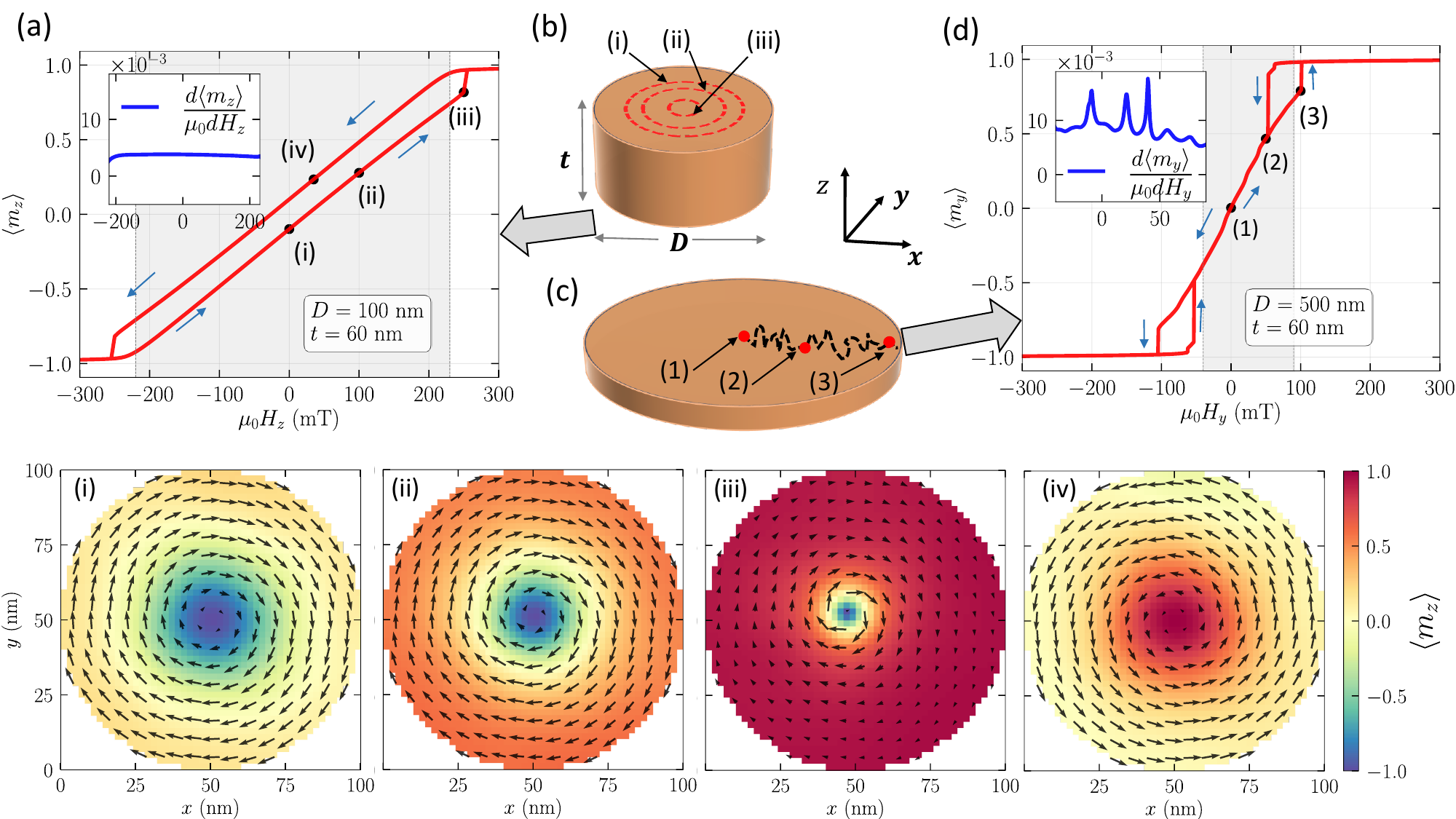}
    \captionsetup{singlelinecheck=false,labelformat=simple, width=\linewidth}
    \caption{  (a) Micromagnetic simulation of a sensor with out-of-plane magnetic field ($H_z$) sensitivity. The simulated transfer curve corresponds to a device with \( D = 100~\text{nm} \) and a sensing layer thickness of \( t = 60~\text{nm} \). The shaded gray region marks the potential dynamic range of the sensor and the interval used for derivative calculations. Snapshots of the magnetic texture at representative points along the hysteresis loop—labeled (i)–(iv)—are shown below the curve, where the arrows indicate the in-plane curling of magnetic moments, while the color scale represents their out-of-plane components, illustrating the presence and progressive evolution of the vortex core with magnetic field. The inset shows the derivative of one branch of the transfer curve, demonstrating the suppression of Barkhausen-type noise. 
(b) Schematic illustration of vortex core contraction under increasing applied magnetic field. 
(c) Schematic of vortex core motion in a conventional in-plane vortex sensor, where an applied magnetic field ($H_y$) induces lateral displacement of the vortex core along the $x$-direction. The intermittent trapping and detrapping of the core at local pinning sites generate Barkhausen-type noise.
(d) Micromagnetic simulation of a noisy transfer curve for an in-plane vortex sensor ($D = 100~\text{nm}$, $t = 60~\text{nm}$) The inset shows the derivative of the transfer curve, highlighting the presence of Barkhausen-type noise. A detailed vortex dynamics of in-plane sensor is shown in Figure S1 of Supporting Information.
}
    \label{fig:figure_2}
\end{figure*}

\autoref{fig:figure_2} presents a micromagnetic simulation study detailing the magnetization evolution of a device with a diameter $D=100$ and thickness $t=60$ nm under an applied out-of-plane magnetic field ($H_z$). The resultant transfer curve in \autoref{fig:figure_2}a exhibits a wide dynamic range exceeding 400 mT. The evolution of the vortex core at selected points along the curve, labeled (i), (ii) and (iii), show progressive shrinking of the core with increasing magnetic field. This process eventually leads to a saturated magnetic state once the field exceeds the annihilation threshold. Upon decreasing the magnetic field from saturation, the transfer curve exhibits hysteresis due to vortex core reversal during the re-nucleation process, as indicated by point~(iv) on the curve. The inset of \autoref{fig:figure_2}a displays the derivative of a single branch of the transfer curve. The smooth profile of this derivative indicates the absence of discontinuities, suggesting a negligible presence of Barkhausen-type noise. This behavior stands in clear contrast to the discrete, jump-like fluctuations commonly observed in conventional in-plane vortex-based sensors~\cite{jotta2021spin, guslienko2001field}, which arise from interactions between the vortex core and pinning centers during its motion across the sample. A representative micromagnetic simulation of an in-plane vortex-based sensor with \( D = 500 \,\mathrm{nm} \) and \( t = 60 \,\mathrm{nm} \) is shown in \autoref{fig:figure_2}d. It is evident that such a sensor exhibits a relatively limited dynamic range ($<$100\,mT). Moreover, defect-induced Barkhausen-type noise is prevalent in this configuration, as illustrated in the inset of \autoref{fig:figure_2}d. For additional context, corresponding micromagnetic simulations of in-plane vortex dynamics are provided in Figure~S1 of the Supporting Information.

The presence of disorder within the free layer plays a critical role in shaping key sensor performance metrics, including the nucleation field, the annihilation field, and the sensitivity (see Figure~S2 of the Supporting Information). Structural or magnetic inhomogeneities increase both the nucleation and annihilation fields, thereby extending the effective operating range of the device. This enhancement, however, is accompanied by a reduction in sensitivity. Furthermore, disorder leads to smoother magnetization transitions near the nucleation and annihilation thresholds, in contrast to the abrupt switching behavior characteristic of ideal, defect-free systems. This trend is in agreement with experimental observations, as illustrated in \autoref{fig:figure_1}d–e.

\begin{figure}[!ht]
\centering
    \includegraphics[width=0.40\linewidth]{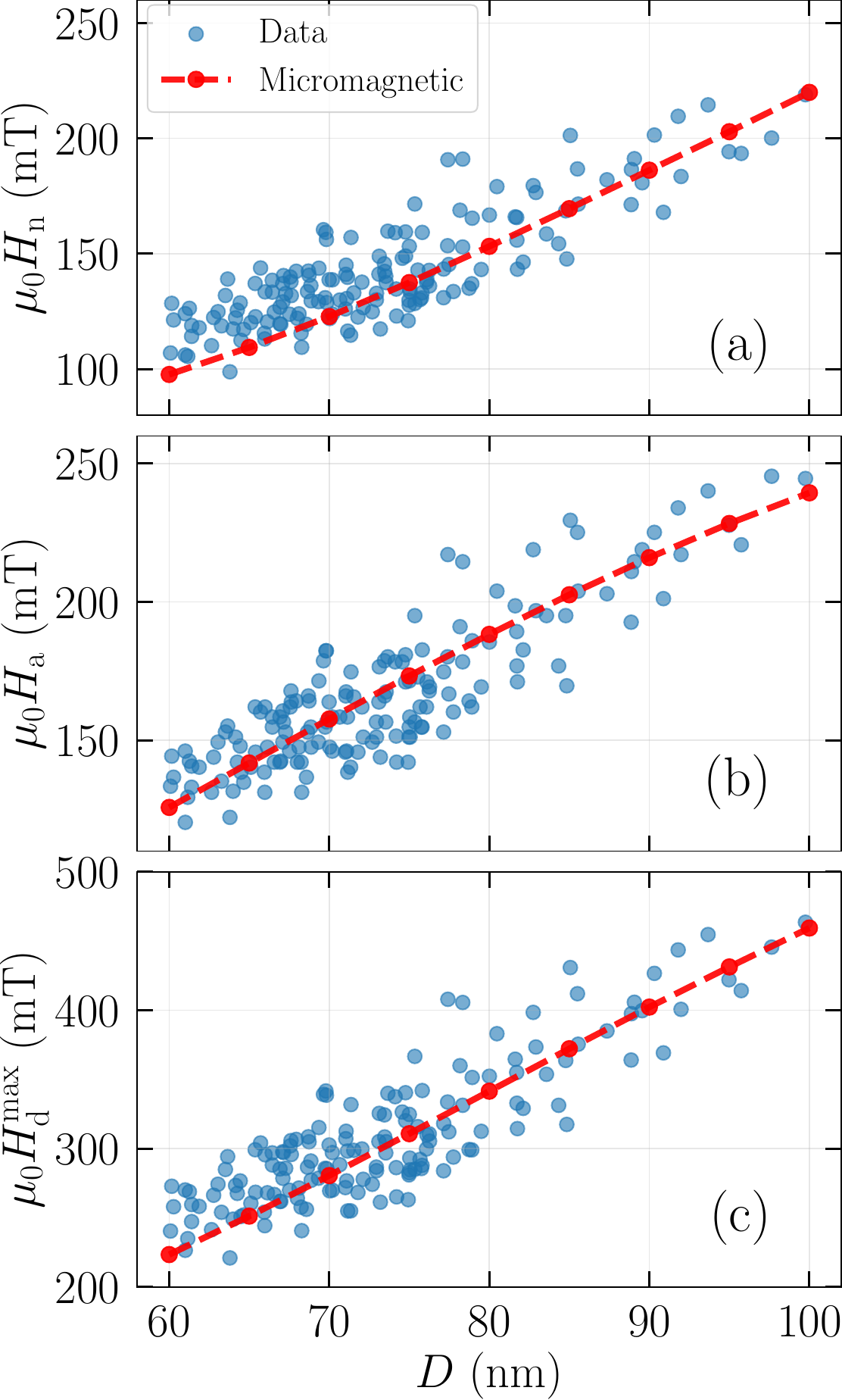}
    \captionsetup{singlelinecheck=false,labelformat=simple, width=\linewidth}
   \caption{(a–c) Experimentally measured nucleation field (\(H_\mathrm{n}\)), annihilation field (\(H_\mathrm{a}\)), and maximum dynamic range (\(H_\mathrm{d}^{\mathrm{max}}\)) for devices with varying diameters (blue scatter points), shown alongside corresponding micromagnetic simulation results (red dashed lines). In all cases, the sensing layer thickness is fixed at 57 nm.
   }
    \label{fig:figure_3}
\end{figure}
  
To further investigate the sensor behavior, experimentally extracted critical parameters such as the nucleation field (\( H_\mathrm{n} \)), annihilation field (\( H_\mathrm{a} \)), and maximum dynamic range \(( H_\mathrm{d}^{\mathrm{max}} = H_\mathrm{a} - H_\mathrm{n} )\) are plotted as a function of device diameter in \autoref{fig:figure_3}. The micromagnetic simulation results show excellent agreement with the experimental data, validating the accuracy of the modeling approach and demonstrating strong consistency with experimental trends. The data reveal that \( H_\mathrm{n} \), \( H_\mathrm{a} \), and \( H_\mathrm{d}^{\mathrm{max}} \) increase approximately linearly with diameter, with \( \mu_0 H_\mathrm{d}^{\mathrm{max}} \) exceeding 450\,mT for a 100\,nm device. These results highlight the exceptional dynamic range achievable in these nanoscale vortex-based sensors.

The dispersion associated with the critical field measurements are attributed to variability of the device shape, which deviates from ideal cylindrical geometry due to nanofabrication limitations. Furthermore, device diameters were estimated based on electrical resistance measurements, assuming a nominal resistance-area product of $10\,\Omega\cdot\mu\text{m}^2$. However, series resistance contributions arising from nanofabrication process variability and partial shunting of the tunnel barrier, by redeposition during pillar definition impact measured resistance values, leading to deviations in the calculated electrical diameters. Both will contribute to the dispersion in calculated physical dimensions and associated critical parameters. Nevertheless, the median values of the experimental data align remarkably well with the micromagnetic simulation results, reinforcing the predictive accuracy of the simulation strategy proposed for optimizing the performance metrics of vortex-based MTJ sensors.

\subsection*{Noise measurements}
Noise characterization is a critical aspect in the evaluation of the performance of MTJ-based sensors, particularly for applications requiring high sensitivity and signal accuracy~\cite{freitas2007magnetoresistive}. Noise in MTJs arises from a variety of mechanisms, including electrical and magnetic thermal noise, shot noise, electronic and magnetic $1/f$ noise, and random telegraph noise (RTN). These noise sources have already been extensively studied and characterized in existing literature~\cite{lei2011review, weitensfelder2018comparison, freitas2007magnetoresistive}. 
The voltage noise spectral density \( S_V \), in the absence of RTN, can be expressed as~\cite{lei2011review, weitensfelder2018comparison}:
\begin{equation}
S_V = 2eV_{\text{bias}}R\cdot\coth\left(\frac{eV_{\text{bias}}}{2N k_B T}\right) + \frac{\alpha_H}{NA} \frac{V_{\text{bias}}^2}{f} \quad \left[\frac{\text{V}^2}{\text{Hz}}\right],
\label{eq:voltage_noise_spectrum}
\end{equation}
where \( e \) is the elementary charge, \( k_B \) is the Boltzmann constant, and \( T \) is the absolute temperature. \( R \) denotes the resistance of a single MTJ sensor, and \( V_{\text{bias}} \) is the total applied voltage in the sensor array. \( N \) represents the number of sensors connected in series, \( \alpha_H \) is the Hooge parameter, \( A \) is the junction area, and \( f \) is the frequency at which the noise is measured.
The first term in \autoref{eq:voltage_noise_spectrum} represents the combined thermal and shot noise contribution. In the limit \( eV_{\text{bias}} \ll k_B T \), this term reduces to the classical Johnson-Nyquist thermal noise \( 4Nk_B T R \)~\cite{johnson1928thermal, nyquist1928thermal}. In contrast, when \( eV_{\text{bias}} \gg k_B T \), it approaches the expression of the shot noise \( 2eV_{\text{bias}} R \)~\cite{blanter2000shot}. These noise sources dominate at high frequencies and originate from distinct physical mechanisms: thermal noise arises from the random motion of charge carriers, whereas shot noise results from the discrete nature of electron tunneling across the junction. Thermal noise contributes a frequency-independent component to the spectrum, leading to a flat noise floor.
The second term in \autoref{eq:voltage_noise_spectrum} corresponds to the \( 1/f \) noise, modeled here using a Hooge-type empirical expression~\cite{hooge20021}. Unlike the original Hooge formulation, which is based on the number of charge carriers, this expression includes a junction area dependence, more appropriate for MTJs, where \( 1/f \) noise has been experimentally observed to scale inversely with the surface area~\cite{ lei2011review, suess2018topologically}. This low-frequency noise component is typically attributed to charge trapping and detrapping processes within the tunnel barrier (electrical \( 1/f \) noise), as well as thermally activated magnetization fluctuations in the free layer, including domain-wall motion and ripple-domain dynamics (magnetic \( 1/f \) noise). The Hooge parameter, $\alpha_H$, is a widely used phenomenological constant to quantify $1/f$ noise. It provides a normalized measure of noise for meaningful comparisons between devices with similar resistance–area (RA) products and magnetic volumes~\cite{weitensfelder2018comparison, lei2011review, suess2018topologically, monteblanco2021normalization}. 

Noise characterization of the vortex-based MTJ sensors was carried out on a single junction, using a battery-powered source to bias the sensor and thereby minimize the influence of external noise sources. The output voltage signal was amplified by a factor of 5000 using a Stanford Research Systems SR560 low-noise voltage preamplifier. A bandpass filter ranging from 0.1~Hz to 10~kHz was applied to suppress the DC component and isolate the relevant frequency band. The amplified signal was then analyzed using a SR780 spectrum analyzer to extract the voltage noise power spectral density, \( \sqrt{S_V} \), via a fast Fourier transform.

Another key performance parameter of a sensor is the sensitivity, which reflects the capability of the sensor to detect small variations in the applied magnetic field. For MTJ sensors, this corresponds to the change in electrical resistance in response to an external magnetic field. The normalized sensitivity, \( \gamma_R \), is defined as:
\begin{equation}
    \gamma_R = \frac{1}{\mu_0R_0} \frac{\mathrm{d}R}{\mathrm{d}H}\cdot 100\quad     \approx \frac{2 \cdot \mathrm{TMR}}{(\mathrm{TMR} + 2) \mu_0H_d^{\mathrm{max}}}\cdot 100\quad
    \left[\frac{\%}{\text{T}}\right],
    \label{eq:sensitivity}
\end{equation}
where \( R_0 \) and \( \mu_0 \) denote the resistance at zero applied magnetic field and the permeability of free space, respectively. TMR is defined as \( \text{TMR} = (R_{AP} - R_P)/R_P \), where \( R_P \) and \( R_{AP} \) are the resistances in the parallel and antiparallel magnetic configurations. $H_d^{\mathrm{max}}$ denotes the maximum dynamic range, which is used here to define the sensitivity in full-scale TMR.

However, in this work, an alternative method used to determine $\gamma_R$ involves applying a small calibrated oscillatory magnetic field with an RMS amplitude of $\mu_0 H_{\text{rms}} = 3~\text{mT}$ at a frequency of $5~\text{Hz}$. The resulting voltage signal exhibits a distinct peak at 5~Hz on frequency-dependent noise spectrum, with an RMS amplitude denoted as \( V_{5\,\text{Hz}}^{\text{rms}} \). Using this value, the sensor sensitivity is calculated as:

\begin{equation}
    \gamma_R = \frac{V_{5\,\text{Hz}}^{\text{rms}}}{\mu_0H_{\text{rms}} \cdot V_{\text{bias}}} \cdot 100 \quad \left[\frac{\%}{\text{T}}\right],
    \label{eq:sensitivity_2}
\end{equation}

where \( V_{\text{bias}} \) is the DC bias voltage applied across the sensor.

Detectivity is a key performance metric that represents the smallest magnetic field the sensor can reliably detect in the presence of noise. A lower value of detectivity corresponds to higher sensitivity and improved noise performance, thereby indicating better sensor quality. Detectivity ($D_T$) was calculated using the following expression:
\begin{equation}
D_T = \frac{\sqrt{S_V}}{\gamma_R \ V_{\text{bias}}} \quad \left[\frac{\text{T}}{\sqrt{\text{Hz}}}\right].
\label{eq:detectivity}
\end{equation}

\begin{figure*}[!htb]
    \includegraphics[width=\linewidth]{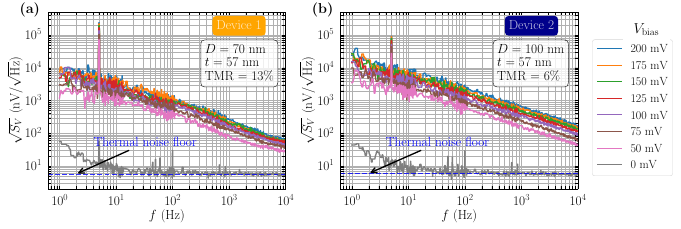}
    \captionsetup{singlelinecheck=false,labelformat=simple, width=\linewidth}
    \caption{
    (a) Frequency-dependent noise measurements for Device~1 and (b) Device~2 under various applied bias voltages. At zero bias, the measured spectra correspond to the background and thermal noise floor. A distinct peak at 5 Hz is observed in both devices, originating from the auxiliary oscillating magnetic field used to probe real-time sensitivity. The noise spectra show typical $1/f$ behavior at low frequencies, which increases with bias.
    }
	\label{fig:figure_4}
\end{figure*}

Frequency-dependent noise measurements were carried out on two devices: Device~1 (\( D = 70\,\mathrm{nm} \), TMR = 13\%) and Device~2 (\( D = 100\,\mathrm{nm} \), TMR = 6\%), under increasing applied voltage bias (see \autoref{fig:figure_4}). The recorded spectra exhibit typical noise behavior of MTJ-based sensors. At all bias conditions, the measured noise levels significantly exceed the ambient background noise, measured at zero bias,  confirming their intrinsic origin. The spectra are primarily dominated by $1/f$-type noise, with the noise magnitude systematically increasing with the applied bias voltage.
Additionally, distinct peaks observed at 5 Hz in each spectrum correspond to the calibrated oscillatory magnetic field applied during the noise measurement. These peaks are used to extract the sensitivity of each sensor using \autoref{eq:sensitivity_2}.

\begin{figure*}[!ht]
\centering
    \includegraphics[width=0.90\linewidth]{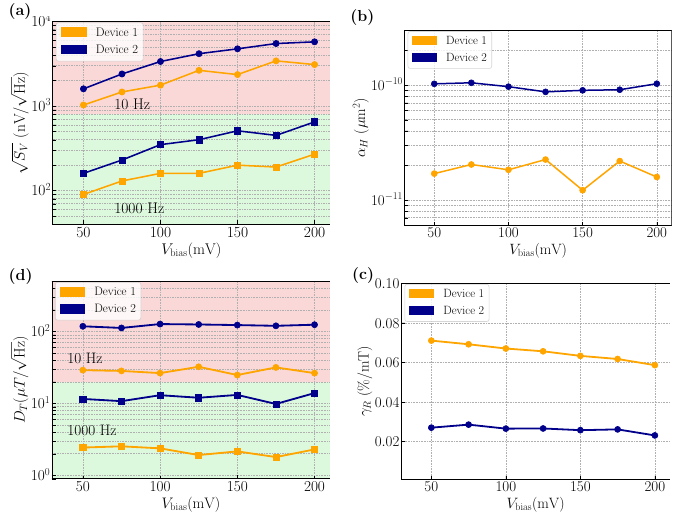}
    \captionsetup{singlelinecheck=false,labelformat=simple, width=\linewidth}
    \caption{  Sensor performance metrics for Device 1 and Device 2.
    (a) Voltage noise spectral density (\(\sqrt{S_V}\)) of the sensor as a function of applied bias at two frequencies: 10~Hz (pink region) and 1000~Hz (green region). 
    (b) Hooge parameter (\( \alpha_H \)) as a function of applied bias. 
    (c) Sensitivity (\( \gamma_R \)) as a function of bias voltage. 
    (d) Detectivity (\( D_T \)) as a function of applied bias for both devices at 10~Hz (pink region) and 1000~Hz (green region).
    }
	\label{fig:figure_5}
\end{figure*}

\autoref{fig:figure_5}a shows the voltage noise spectral density (\( \sqrt{S_V} \)) for both sensors at 10\,Hz and 1000\,Hz as a function of the applied bias voltage. In both cases, the noise increases monotonically with increasing bias. In particular, the noise at 1000\,Hz is at least one order of magnitude lower than at 10\,Hz for both devices. Furthermore, Device~1 consistently exhibits lower noise than Device~2, despite the latter having a larger area. For example, at 1000\,Hz and at 100\,mV bias, Device~1 exhibits a noise level of approximately 160\,nV/\( \sqrt{\text{Hz}} \), whereas Device~2 reaches around 350\,nV/\( \sqrt{\text{Hz}} \). Since the noise of a sensor is inversely proportional to its active area, appropriate area normalization is required to obtain an intrinsic noise parameter. This normalization enables meaningful performance comparisons between devices of different sizes. To achieve this, each noise spectrum was fitted using \autoref{eq:voltage_noise_spectrum}.
Figures~S3 in the Supporting Information display the measured noise spectra along with their corresponding fitted curves for both devices, demonstrating a strong agreement between the experimental results and the theoretical model. The extracted Hooge parameter, \( \alpha_H \), derived from these fits, is plotted in \autoref{fig:figure_5}b. Notably, \( \alpha_H \) remains approximately constant throughout the bias range, with average values of approximately \( 2 \times 10^{-11}~\mu\text{m}^2 \) for Device~1 and \( 1 \times 10^{-10}~\mu\text{m}^2 \) for Device~2. These results underscore a clear correlation between reduced noise and higher TMR ratios~\cite{lei2011review}, likely attributable to a lower density of intrinsic and extrinsic defects in a device with higher TMR. Furthermore, the Hooge parameter observed in the present vortex-based sensors are lower than previously reported for conventional MTJ-based vortex sensors~\cite{he2020nonhysteretic, weitensfelder2018comparison, suess2018topologically}, indicating the intrinsic noise advantage of the proposed design.

\definecolor{LightRed}{RGB}{255, 230, 230}
\definecolor{LightBlue}{RGB}{230, 240, 255}
\definecolor{LightGray}{gray}{0.98}
\definecolor{orange}{RGB}{255, 165, 0}
\definecolor{LightOrange}{RGB}{255, 190, 100}
\definecolor{MidDarkBlue}{RGB}{150, 180, 220}
\definecolor{LightViolet}{RGB}{230, 230, 255}
\begin{table*}[!tbp]
\centering
\caption{ 
Performance metrics comparison of vortex-based magnetic sensors, including previously reported devices and the proposed sensors (Device~1 and Device~2). ``Field sense'' indicates the orientation of the magnetic field to which the sensor is sensitive: in-plane (IP) or out-of-plane (OOP). \( D \) and \( t \) denote the diameter and thickness of the sensing layer, respectively, and \( N \) represents the number of sensors in an array. \( H_\mathrm{d} \) refers to the dynamic range, \( \sqrt{S_V} \) is the voltage noise spectral density, and \( D_T \) denotes the detectivity. \( \alpha_H \) represents the Hooge parameter, and \( \gamma_R \) is the sensitivity. The figure of merit (FOM) (see \autoref{eq:FOM})  is also listed, where lower values correspond to higher resolution and improved overall sensor performance. 
The parameters \( \sqrt{S_V} \), \( D_T \), and FOM are reported in two forms: (i) as originally measured for arrays containing \( N \) elements, denoted as ``Rep,'' and (ii) as area-normalized values, assuming a sensing area equivalent to that of Device~1, denoted as ``Norm.'' This normalization enables a fair comparison of performance metrics across devices of different sizes. The final row presents the estimated performance of the proposed sensor under full-scale normalization (TMR = 100\%), extrapolated from experimental trends.
}
\label{tab:table_1}
\renewcommand{\arraystretch}{2}
\setlength{\tabcolsep}{3pt}
\resizebox{\textwidth}{!}{%
\begin{tabular}{|c|c|c|c|c|c|c|c|
c|
c|c|c|c|c|c|}
\hline
\hline
\multirow{2}{*}{\textbf{Device}} &
\multirow{2}{*}{\makecell{\textbf{Field} \\ \textbf{Sense}}} &
\multirow{2}{*}{\makecell{$\boldsymbol{D}$ \\[1.5ex](nm)}} &
\multirow{2}{*}{\makecell{$\boldsymbol{t}$ \\[1.5ex](nm)}} &
\multirow{2}{*}{\makecell{$\boldsymbol{N}$ \\[4ex]} } &
\multirow{2}{*}{\makecell{\textbf{TMR} \\[1.5ex](\%)}} &
\multirow{2}{*}{\makecell{$\boldsymbol{H_d}$ \\[1.5ex](mT)}} &
\multicolumn{2}{c|}{\makecell{$\boldsymbol{\sqrt{S_V}}$ (nV/$\sqrt{\mathrm{Hz}}$)\\@10 Hz}} &
\multirow{2}{*}{\makecell{$\boldsymbol{\alpha_H}$ \\[1.5ex]($\mu$m$^2$)}} &
\multirow{2}{*}{\makecell{$\boldsymbol{\gamma_R}$ \\[1.5ex](\%/mT)}} &
\multicolumn{2}{c|}{\makecell{$\boldsymbol{D_T}$ (nT/$\sqrt{\mathrm{Hz}}$)\\@10 Hz}} &
\multicolumn{2}{c|}{\makecell{\textbf{FOM} (ppm/$\sqrt{\mathrm{Hz}}$)\\@10 Hz}} \\
\cline{8-9} \cline{12-13}\cline{14-15}
& & & & & & & \makecell{Rep.\textsuperscript{\dag}} & \makecell{Norm.\textsuperscript{\textsection}} & & & \makecell{Rep.\textsuperscript{\dag}} & \makecell{Norm.\textsuperscript{\textsection}} & 
\makecell{Rep.\textsuperscript{\dag}} & \makecell{Norm.\textsuperscript{\textsection}} \\
\hline
\rowcolor{LightViolet}
Vortex GMR~\cite{suess2018topologically} & IP & 2000 & 70 & 1752 & 5.6 & 80 & 20 & 23918 & n.a. & 0.04 & 20 & 23918 & 0.25. & 299  \\
\rowcolor{gray!30}
Vortex TMR~\cite{weitensfelder2018comparison} & IP & 2000 & 80 & 10 & 85 & $\sim 60$ & 200 & 18070 & $\sim 10^{-9}$ & 0.6 & 300 & 27105 & 5 &  452 \\
\rowcolor{LightOrange}
Vortex Device 1 & OOP & 70 & 57 & 1 & 13 & 300 & 1770 & 1770 & $2 \times 10^{-11}$ & 0.07 & 26430 & 26430 & 88 &  88 \\
\rowcolor{MidDarkBlue}
Vortex Device 2 & OOP & 100 & 57 & 1 & 6 & 400 & 3370 & 4814 & $1 \times 10^{-10}$ & 0.03 & 126930 & 181329 & 317 &  453 \\
 \rowcolor{LightRed}
Vortex Device 1\textsuperscript{\#} & OOP & 70 & 57 & 1 & 100 & 300 & 1612 & 1612 & $\sim 10^{-11}$ & 0.22 & 7327 & 7327 & 24 & 24  \\
\hline
\hline
\end{tabular}
}
\raggedright
\footnotesize{
\hspace{0.5cm}
\textsuperscript{\dag} Reported/measured across $N$ nominally identical devices arranged in an array.\\
\textsuperscript{\textsection} Normalized to single-junction area of Device 1.\\
\textsuperscript{\#} An estimated device with full scale normalization (TMR = 100\%).
}
\end{table*}
\autoref{fig:figure_5}c shows the normalized sensitivity of the sensors, calculated using \autoref{eq:sensitivity_2}. The results indicate that the sensor with a higher TMR exhibits greater sensitivity and, as expected, enhancing the TMR is an effective strategy to improve sensor performance. Therefore, to evaluate the full potential of the sensor, Device 1 is also reported using full-scale normalization, which is equivalent to assigning a 100\% TMR signal (see \autoref{tab:table_1}). Such a response should be achievable in these junctions~\cite{sidi2022size}. Therefore, according to \autoref{eq:sensitivity}, a dynamic range of 300\,mT corresponds to a sensitivity of 0.22\%/mT, whereas a reduced dynamic range of 200\,mT would yield a higher sensitivity of 0.33\%/mT. These values are particularly notable for sensors operating over such wide dynamic ranges, while maintaining sensitivities comparable to those of Hall sensors~\cite{ripka2010advances, khan2021magnetic}. 

A comparison of the detectivity \( D_T \), calculated using \autoref{eq:detectivity}, for Device~1 and Device~2 at 10~Hz and 1000~Hz is shown in \autoref{fig:figure_5}d. The plot indicates that the detectivity of both devices remains nearly constant throughout the applied bias voltage range. Furthermore, the device with the higher TMR (Device~1) consistently exhibits superior detectivity compared to the device with the lower TMR (Device~2).
Moreover, extending the preceding analysis and assuming a TMR ratio of 100\% along with a dynamic range of 300\,mT, the projected detectivity is approximately 7327\,nT/\( \sqrt{\text{Hz}} \) at 10\,Hz and 733\,nT/\( \sqrt{\text{Hz}} \) at 1000\,Hz. These values represent a significant performance achievement for a single MTJ-based sensor, highlighting the advantage of vortex-based nanoscale magnetic field sensors for applications requiring a broad field range with competitive sensitivity and detectivity.

\autoref{tab:table_1} compares the key performance parameters of the sensor presented with those of previously reported vortex-based sensors. For example, a vortex GMR sensor sensitive to in-plane magnetic fields~\cite{suess2018topologically}, comprising an array of 1752 sensor elements, has been reported to exhibit a noise level of 20\,nV/\(\sqrt{\mathrm{Hz}}\). However, each individual sensor has a diameter of 2\,\(\mu\)m, resulting in a total sensing area of \(N A_{\mathrm{MTJ}} = 5.5 \times 10^{9}\mathrm{nm}^2\). For a fair comparison with the sensor presented here, the performance metrics are normalized to an equivalent sensing area. In this work, Vortex Device 1 (\(D = 70\,\mathrm{nm}\)) with an active area of 3848.45 nm\(^2\) is used as a reference. The corresponding normalization factor is given by \(\sqrt{\frac{5.5 \times 10^{9}\mathrm{nm}^2}{3848.45~\mathrm{nm}^2}} \approx 1195\). After applying this normalization, the noise at 10\,Hz for the GMR vortex sensor becomes 23,918\,nV/\(\sqrt{\mathrm{Hz}}\), which is 13.5 times higher than that of the vortex sensor presented (1770\,nV /\(\sqrt{\mathrm{Hz}}\)). Following the same procedure, the normalized detectivity is also calculated and listed. In particular, despite the vortex device in this work exhibiting a much larger dynamic range (300\,mT) compared to the GMR vortex sensor (80\,mT), the normalized detectivity values remain very similar.
Generally, the detectivity decreases as the dynamic range increases.
For a comparative evaluation of our vortex-based sensor, we compute a figure of merit (FOM), defined in terms of the detectivity and dynamic range, as follows:
\begin{equation}
\mathrm{FOM} = \frac{D_T}{\mu_0 H_\mathrm{d}^{\mathrm{max}}} \times 10^6 
\quad \left(\frac{\mathrm{ppm}}{\sqrt{\mathrm{Hz}}}\right),
\label{eq:FOM}
\end{equation}
where the FOM, expressed in parts per million (ppm), quantifies the resolution of a magnetic field sensor relative to its full-scale range. A lower FOM value corresponds to higher resolution and therefore better sensor performance.

Extending our analysis, the normalized FOM for the vortex GMR sensor is found to be \( 299~\mathrm{ppm}/\sqrt{\mathrm{Hz}} \), corresponding to a resolution of approximately 12-bits. In contrast, the presented vortex Device~1 achieves a FOM of \( 88~\mathrm{ppm}/\sqrt{\mathrm{Hz}} \), equivalent to a resolution of 14-bits. This higher resolution is achieved despite the device operating over a significantly wider dynamic range, underscoring the unique advantage of our nanoscale sensor in accurately measuring magnetic field.

For comparison, we also consider an in-plane vortex TMR sensor~\cite{weitensfelder2018comparison}, which demonstrates inferior performance relative to Device~1 in terms of noise, detectivity, and FOM. However, the performance of Device~2, which has a larger diameter and lower TMR ratio, is degraded (\(\mathrm{FOM} = 453\)) due to the reduced TMR and elevated noise levels, likely arising from higher-order defects in that particular device. This indicates that the performance of the presented sensor can be further enhanced by optimizing the nanofabrication process and the magnetic stack to achieve higher TMR.
For example, if Device~1 (\(D = 70~\mathrm{nm}\)) were improved to achieve a realistic TMR ratio of \(100\%\) with full-scale TMR normalization, a sensitivity as high as \(0.22~\%/\mathrm{mT}\) could be obtained. Consequently, this would yield a detectivity of \(7327~\mathrm{nT}/\sqrt{\mathrm{Hz}}\) at 10~Hz and a FOM of \(24~\mathrm{ppm}/\sqrt{\mathrm{Hz}}\), corresponding to a resolution of approximately 15-bits.

The nanoscale dimensions of the proposed sensor underscore its minimal footprint, where key performance parameters such as noise, detectivity, and FOM can be significantly enhanced through the use of sensor arrays, since these parameters scale as $1/\sqrt{N}$. For instance, an array comprising 1600 nanoscale sensors improves these parameters by a factor of 40, yielding an effective noise level of $40.3~\text{nV}/\sqrt{\text{Hz}}$, a detectivity of $183.7~\text{nT}/\sqrt{\text{Hz}}$, and an FOM of $0.6~\text{ppm}/\sqrt{\text{Hz}}$, all calculated at 10~Hz, corresponding to a 21-bit resolution.  

From an area perspective, an array design with a sensor diameter of 100~nm and a pitch of 400~nm, arranged in a $40 \times 40$ configuration, occupies a total footprint of only $246~\mu\text{m}^2$. By contrast, a conventional in-plane vortex sensor~\cite{suess2018topologically} with a diameter of 2~$\mu$m and a pitch of 8~$\mu$m requires a footprint of $98596~\mu\text{m}^2$ for the same $40 \times 40$ array, corresponding to an approximately 400-fold larger area than the proposed nanoscale sensor architecture.
These results highlight the scalability and performance benefits of the proposed sensor design, particularly for applications requiring a high dynamic range, low noise, and compact form factor.
Figure S4 in the Supporting Information presents the calculated metrics for dynamic range, sensitivity, detectivity, and FOM of the proposed single-element sensor with varying diameters and aspect ratios. These results provide a practical roadmap for tailoring sensor properties through the selection of an appropriate device geometry, thereby enabling performance optimization for specific application requirements.

\section{Conclusion} 
\label{sec:conclusion}

In this study, we have demonstrated a nanoscale vortex-based MTJ sensor optimized for the detection of perpendicular magnetic fields. The sensor achieves a dynamic range exceeding 200\,mT while maintaining low intrinsic noise, attributed to the vortex-state magnetization dynamics involving the expansion and contraction of the vortex core, inherently reducing noise arising from local defects. Micromagnetic simulations incorporating defects within the sensing layer accurately reproduce the observed sensor behavior, providing a reliable modeling benchmark for predicting device performance.

Frequency-dependent noise measurements reveal that the sensor is dominated by \( 1/f \)-type noise, with low Hooge parameters (\( 2 \times 10^{-11}~\mu\mathrm{m}^2 \)) and detectivity values comparable to those of previously reported in-plane vortex sensors. With full-scale normalization (TMR = 100\%), the proposed sensor can achieve detectivity values of 7327\,nT/\(\sqrt{\mathrm{Hz}}\) at 10\,Hz and 733\,nT/\(\sqrt{\mathrm{Hz}}\) at 1000\,Hz for a single junction---remarkably low for a single MTJ element operating over such a broad dynamic range. Moreover, the figure of merit indicates a higher bit resolution (14-bit) compared to conventional in-plane vortex sensors (12-bit), surpassing previously reported values for MTJ-based sensors, which typically operate over a narrower dynamic range of 40--80\,mT. This demonstrates the excellent balance achieved between sensitivity, noise performance, and dynamic range in a device with a lateral dimension below 100\,nm.
The nanoscale footprint of the sensor further enables array-based architectures, with performance improvements expected to scale as \( 1/\sqrt{N} \), opening pathways for applications requiring large dynamic range, low noise, high resolution and minimal footprint.

Thus, the proposed vortex-based MTJ architecture offers a promising benchmark for high-performance, wide-field-range magnetic sensing, providing a scalable and integrable platform for TMR magnetic field-sensing applications.

\section*{Methods}
\hypertarget{sec:methods}{}
\subsubsection{Sample preparation}
The samples were fabricated by magnetron sputtering at an argon pressure of \(2 \times 10^{-3}\,\text{mbar}\) on a 100\,mm silicon wafer. The complete magnetic stack is made of:
\noindent
seed layers/\allowbreak[Co(0.5)/\allowbreak Pt(0.25)]$_{\times 6}$\allowbreak
/\allowbreak Co(0.5)\allowbreak
/\allowbreak Ru(0.9)\allowbreak
/\allowbreak[Co(0.5)/\allowbreak Pt(0.25)]$_{\times 3}$\allowbreak
/\allowbreak W(0.2)\allowbreak
/\allowbreak FeCoB(1)\allowbreak
/\allowbreak MgO(1.1)\allowbreak
/\allowbreak FeCoB(1.4)\allowbreak
/\allowbreak W(0.2)\allowbreak
/\allowbreak NiFe(57)\allowbreak
/\allowbreak Ta(1)\allowbreak
/\allowbreak Pt(3),  where the values in parentheses denote layer thicknesses in nanometers. The lower part of the stack incorporates a synthetic antiferromagnetic (SAF) structure formed by
[Co(0.5)/Pt(0.25)]$_{\times 6}$\allowbreak/Co(0.5)\allowbreak/Ru(0.9)\allowbreak/[Co(0.5)/Pt(0.25)]$_{\times 3}$
which is exchange-coupled to the FeCoB(1) reference layer (pinned).

\noindent
The MgO tunnel barrier was made through a two-step process involving the deposition of a metallic Mg layer, followed by oxidation at a pressure of \(10^{-2}\,\text{mbar}\) for 10 seconds and subsequently capped with an additional Mg layer. Above the tunnel barrier, a composite sensing layer consisting of FeCoB(1.4)/W(0.2)/NiFe(57) was deposited.
A protective capping layer of Ta(1)/Pt(3) was added to prevent ambient oxidation.
Post-deposition annealing was carried out at \(300\,^\circ\mathrm{C}\) for 10 minutes under high vacuum conditions (\(5 \times 10^{-6}\,\text{mbar}\)) to promote the crystallization of the MgO barrier and enhance the PMA in the FeCoB layers. The resistance area (RA) product, measured using the current-in-plane tunneling (CIPT) method, was found to be \(10\,\Omega\cdot\mu\text{m}^2\). Subsequently, nanopillar devices with nominal diameters ranging from 60\,nm to 100\,nm were defined using electron beam lithography, ion beam etching, and standard photolithography followed by metallization processes.

\subsubsection{Micromagnetic model}
Micromagnetic simulations were performed using the GPU-accelerated MuMAX software package$^3$ \cite{vansteenkiste2014design}. The simulations were carried out on sensing layers with varying diameter and thickness. The computational domain was discretized using a finite-difference scheme with a uniform cell size of \(3 \times 3 \times 3\,\text{nm}^3\), providing an optimal trade-off between computational accuracy and time efficiency. In the simulations, standard material parameters representative of permalloy (NiFe) were employed: saturation magnetization \( M_s = 8 \times 10^5\,\mathrm{A/m} \), exchange stiffness constant \( A = 13 \times 10^{-12}\,\mathrm{J/m} \), and Gilbert damping factor \( \alpha = 0.01 \).

To incorporate the influence of disorder and pinning potentials, factors that are intrinsic to deposited polycrystalline thin films, the simulation geometry was modified using a Voronoi tessellation approach. This method subdivides the magnetic structure into randomly distributed grains, with an average grain size of around \SI{20}{\nano\meter}, thereby emulating the polycrystalline morphology of the sensing layer. To further capture the influence of disorder, the exchange stiffness at grain boundaries was reduced by 10\% relative to the bulk. In addition, interfacial PMA was modeled as a spatially random variable, following a Gaussian distribution with a mean value of \SI{1.4}{\milli\joule\per\meter\squared} and a standard deviation of 20\%. These modifications reflect realistic physical inhomogeneities originating from lattice imperfections, compositional fluctuations, and fabrication-induced defects.

Quasistatic magnetization responses were evaluated by first allowing the system to relax to its equilibrium state. Following this relaxation, the external magnetic field \(B_{\text{ext}}\) was varied in small steps \(\Delta B_{\text{ext}}\) to systematically probe the evolution of the magnetic configuration under near-equilibrium conditions.

\begin{suppinfo}
Micromagnetic simulation of conventional in-plane sensitive sensor; Micromagnetic simulations comparing the effect of disorder on the hysteresis behavior; Noise measurement and fits using $1/f$-type noise;  Micromagnetic simulations of key performance metrics for vortex-based magnetic sensors as a function of diameter and aspect ratio (\( t/D \)). 
\end{suppinfo}

\section{Declarations}
\subsubsection{Author Contributions}
A.J., R.C., L.B.P., and B.D. conceived the idea and planned the study. S.A. prepared the thin-film samples and conducted structural analysis. A.J performed the  magnetic, and transport characterizations. A.J. fabricated the devices and carried out the magnetotransport measurements and analysis. A.J. and H.B. performed the noise measurements and subsequent analysis. A.J. conducted the micromagnetic simulations. R.C., L.B.P., and B.D. coordinated and supervised the project. All authors discussed the results and contributed to writing the manuscript.

\subsubsection{Conflict of interest}
The authors declare no competing interests.

\subsubsection{Availability of data and materials}
The data is available upon reasonable request from corresponding authors.

\begin{acknowledgement}
The authors would like to acknowledge Alvaro Palomino and Nikita Strelkov for their contributions to the preliminary development of the sensing concept. This work was supported partially supported by the European Research Council via grant reference ERC-2022-PoC2 (NANOSENSE No.101100599), French National Research Agency in Project SpinSpike (ANR-20-CE24-0002), `France 2030' program PEPR PRESQUILE (ANR-22-PETQ-0002) and by the French National Research Agency in part through the “France 2030” program PEPR SPIN under Project
ADAGE (ANR-22-EXSP-0006).
\end{acknowledgement}
\bibliography{mybibliography}

\newpage
\begin{center}
	{\large \bfseries Supplementary Information}\\[5ex]
	{\bfseries Low-Noise Nanoscale Vortex Sensor for Out-of-Plane Magnetic Field Detection}\\[1ex]
	Ajay Jha$^{1}$, Alvaro Palomino$^{1}$, Stephane Auffret$^{1}$, Hélène Béa$^{1, 2}$, Ricardo C. Sousa$^{1}$, Liliana D. Buda-Prejbeanu$^{1}$, Bernard Dieny$^{1}$\\[1ex]
	\textit{$^1$Univ. Grenoble Alpes, CEA, CNRS, Grenoble-INP, Spintec, 38000 Grenoble, France}\\
	\textit{$^2$Institut Universitaire de France (IUF), 75000 Paris, France}\\
\end{center}
\vspace{10mm}
\noindent

\setcounter{figure}{0}
\renewcommand{\thesection}{S\arabic{section}}
\renewcommand{\thefigure}{S\arabic{figure}}

\begin{figure*}[!ht]
	\centering
	\includegraphics[width=\linewidth]{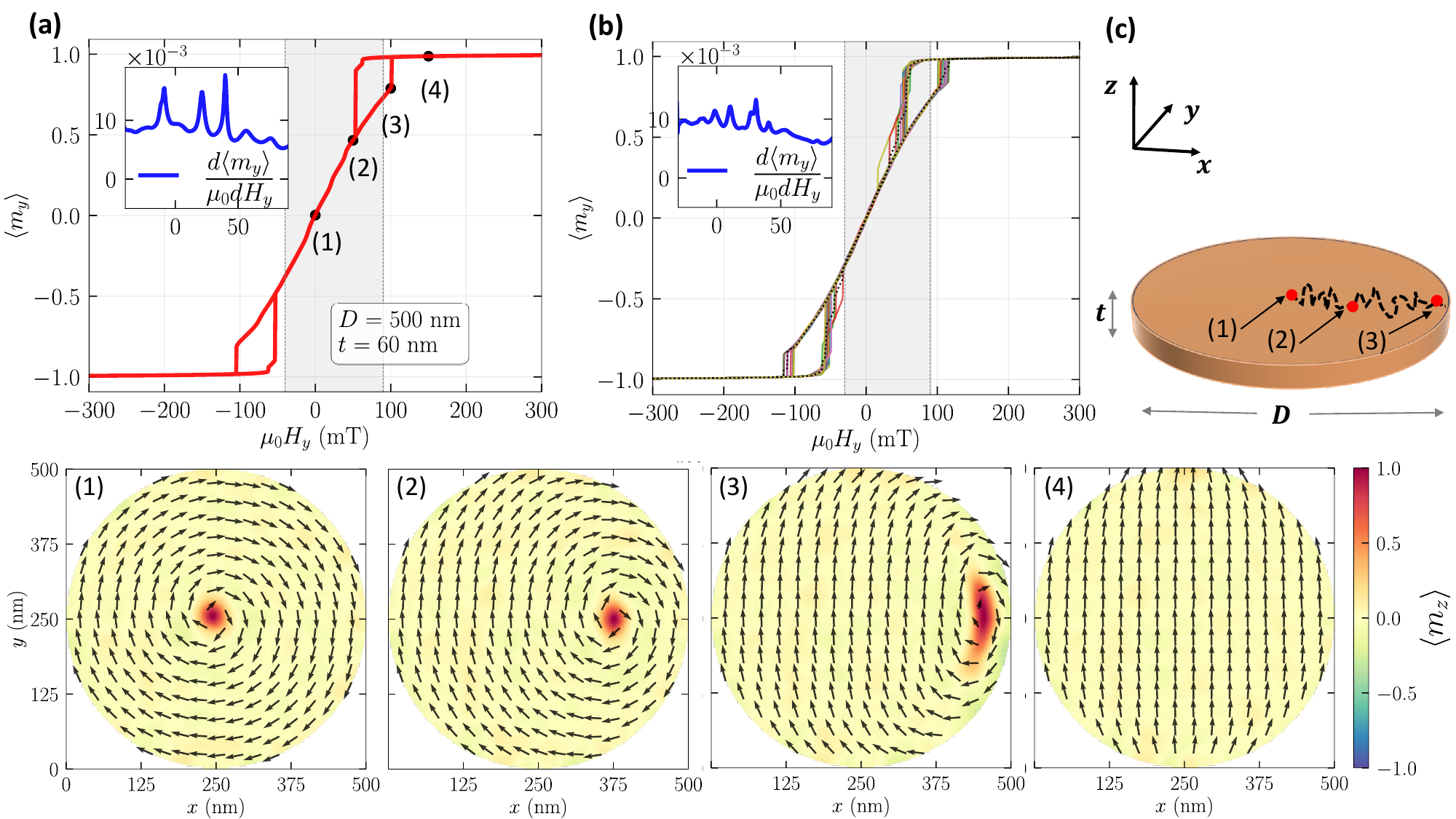}
	\captionsetup{singlelinecheck=false,labelformat=simple, width=\linewidth}
	\caption{Micromagnetic simulation of conventional in-plane sensitive sensor.(a) Simulated transfer curve for a conventional in-plane vortex-based sensor designed for in-plane magnetic field detection. The device has a diameter of \( D = 500~\text{nm} \) and a thickness of \( t = 60~\text{nm} \). Snapshots of the magnetic texture at selected points along the hysteresis loop—labeled (1)-(4) are shown beneath the curve. The shaded gray region marks the potential dynamic range of the sensor and the interval used for derivative calculations. The inset displays the derivative of one branch of the transfer curve, highlighting the presence of prominent Barkhausen-type noise. (b) Simulated transfer curves for an array of 25 such devices, illustrating dispersion in both the nucleation and annihilation fields due to variability in local magnetic pinning. The averaged response demonstrates a partial suppression of noise; however, significant fluctuations remain. (c) Schematic illustration of vortex core dynamics in a conventional in-plane sensor. The applied magnetic field induces lateral motion of the vortex core, which becomes intermittently trapped and detrapped in local pinning sites, giving rise to Barkhausen-type noise.
	}
	\label{fig:sfigure_1}
\end{figure*}

\autoref{fig:sfigure_1}a presents a micromagnetically simulated hysteresis loop for a disc with diameter \( D = 500~\text{nm} \) and thickness \( t = 60~\text{nm} \), subjected to an in-plane magnetic field \( H_y \). Snapshots of the vortex core displacement along the \( x \)-axis, corresponding to selected points on the transfer curve, are also shown in \autoref{fig:sfigure_1}. The operational range of the sensor is indicated by the shaded grey region, representing the magnetic field interval over which the device exhibits a quasi-linear response. The inset displays the derivative of the upper branch of the transfer curve within this region, highlighting the presence of intrinsic step-like features. These discrete fluctuations are attributed to Barkhausen-like noise, originating from pinning and depinning events that occur as the vortex core traverses local energy barriers.

\begin{figure*}[!htb]
	\centering
	\includegraphics[width=0.95\linewidth]{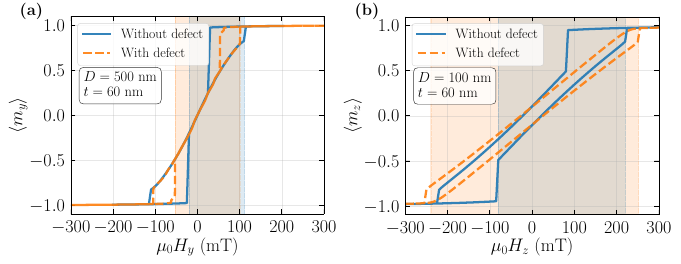}
	\captionsetup{singlelinecheck=false,labelformat=simple, width=\linewidth}
	\caption{Micromagnetic simulations comparing the effect of disorder on the hysteresis behavior of (a) an in-plane vortex sensor with diameter \( D = 500~\text{nm} \) and thickness \( t = 60~\text{nm} \), and (b) an out-of-plane vortex sensor with diameter \( D = 100~\text{nm} \) and thickness \( t = 60~\text{nm} \). For out-of-plane sensor, the introduction of local defects leads to a noticeable shift in the nucleation and annihilation fields, resulting in an extended dynamic range (shaded region) for the sensor.
	}
	\label{fig:sfigure_2}
\end{figure*}

\autoref{fig:sfigure_2} illustrates the influence of structural disorder within the sensing layer for both in-plane and out-of-plane vortex-based sensors. In both configurations, the presence of disorder significantly impacts key performance parameters, including the nucleation field, annihilation field, and sensitivity. Specifically, defects lead to an increase in the critical fields, thereby extending the operational range of sensor. However, this enhancement comes at the expense of reduced sensitivity in both sensing geometries.
Moreover, the presence of disorder induces smoother magnetization transitions near the nucleation and annihilation thresholds, in contrast to the abrupt changes observed under ideal, defect-free conditions. This behavior aligns well with experimental observations, further validating the simulation results presented in the main text (see Figure~1).

\begin{figure}[!htb]
	\centering
	\includegraphics[width=\linewidth]{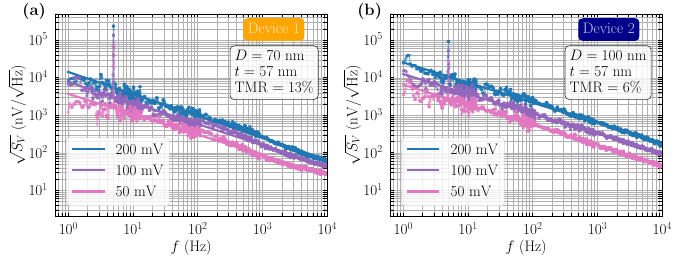}
	\captionsetup{singlelinecheck=false,labelformat=simple, width=\linewidth}
	\caption{Noise measurement and fits using $1/f$-type noise.
		(a) and (b) Voltage noise spectral density (\( \sqrt{S_V}\)) experimental data (dots) for Device~1 and Device~2, respectively, fitted using Equation~1 from the main text (solid lines) for different applied bias voltages.}
	
	\label{fig:sfigure_3}
\end{figure}

\newpage
\autoref{fig:sfigure_3} presents the selected frequency-dependent voltage noise spectral density (\( \sqrt{S_V} \)) data measured at three different bias voltages (50\,mV, 100\,mV, and 200\,mV), along with the corresponding theoretical fits. The spectral density \( \sqrt{S_V} \) is predominantly governed by \( 1/f \)-like noise across the frequency range. The Hooge parameter extracted from these fits provides a normalized noise metric for each device, as discussed in detail in the main text.


\begin{figure*}[!htb]
	\centering
	\includegraphics[width=\linewidth]{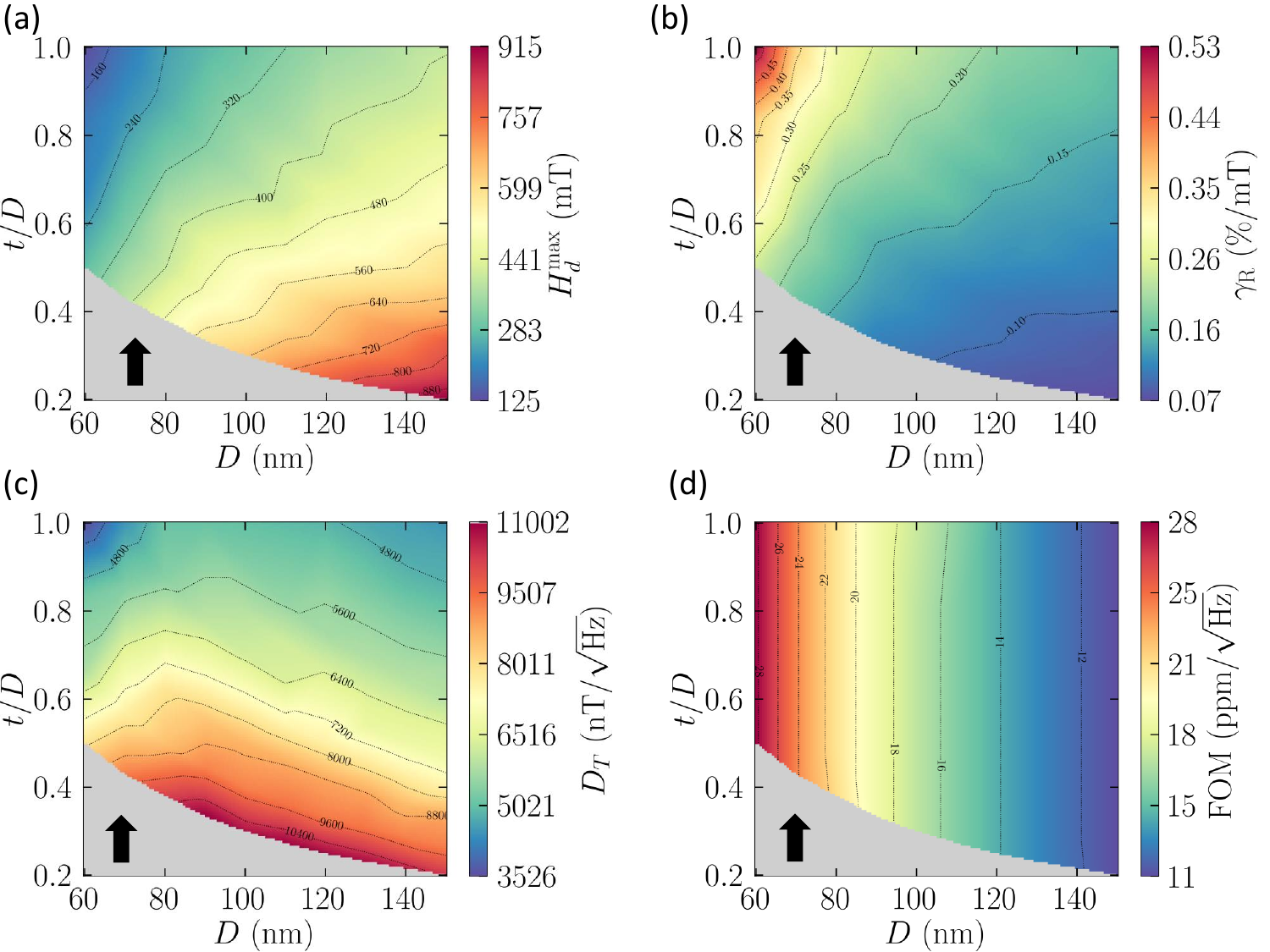}
	\captionsetup{singlelinecheck=false,labelformat=simple, width=\linewidth}
	\caption{
		Micromagnetic simulations of key performance metrics for a vortex-based magnetic sensor as a function of diameter and aspect ratio (\( t/D \)), where \( D \) and \( t \) denote the diameter and thickness of the sensing layer, respectively. The simulated parameters include: (a) maximum dynamic range (\( H_d^\mathrm{max} \)), (b) sensitivity (\( \gamma_R \)), (c) detectivity (\( D \)) and (d) figure of merit (FOM) representation of the sensing layer geometry. These maps enable rapid estimation of sensor performance for a given geometric configuration, or conversely, allow the identification of suitable geometries to meet specific application requirements. The grey shaded region denotes the parameter space where the vortex state does not constitute the ground state; instead, an out-of-plane magnetic configuration becomes stabilized. 
	}
	\label{fig:sfigure_4}
	
\end{figure*}

\autoref{fig:sfigure_4}a presents the simulated dynamic range map of the out-of-plane vortex sensor as a function of device diameter (\( D \)) and aspect ratio (\( t/D \)), where \( t \) is the thickness of the sensing layer. The results indicate that a wide dynamic range exceeding 200\,mT can be achieved for devices with sub-100\,nm diameters. The shaded grey region marks the parameter space in which the vortex state is not stable; instead, the ground state adopts an out-of-plane magnetic configuration due to the dominance of exchange energy over magnetostatic energy.  

Figures~\ref{fig:sfigure_4}b--d display the simulated sensitivity (\( \gamma_R \)), detectivity (\( D_T \)), and figure of merit (FOM) for the proposed device with a single element, based on experimentally derived parameters. Both \( D_T \) and FOM are evaluated for 10\,Hz operation (see Table~1 in the main text). These maps clearly demonstrate that the proposed sensor architecture can be tailored to achieve a wide range of performance metrics, with significantly improved characteristics compared to conventional in-plane vortex sensors.

\end{document}